\documentclass[aps,pra,twocolumn,showpacs,superscriptaddress]{revtex4}
\usepackage{graphicx}
\usepackage{amsmath}
\usepackage{amssymb}

\input{epsf}
\begin{document}

\title{Excitation spectrum and effective interactions
of highly-elongated Fermi gas}

\author{D. Blume}
\affiliation{Department of Physics and Astronomy,
Washington State University,
  Pullman, Washington 99164-2814, USA}
\author{D. Rakshit}
\affiliation{Department of Physics and Astronomy,
Washington State University,
  Pullman, Washington 99164-2814, USA}

\date{\today}

\begin{abstract}
Full 3D calculations of small two-component Fermi gases
under 
highly-elongated confinement, in which unlike
fermions interact through short-range
potentials with variable atom-atom $s$-wave scattering length, 
are performed using the correlated Gaussian approach. 
In addition, microscopic 1D calculations 
are performed for 
effective ``atomic'' and ``molecular'' 1D model Hamiltonian.
Comparisons of the 3D and 1D energies and 
excitation frequencies 
establish the validity regimes of the effective 1D Hamiltonian.
Our numerical results for three- and four-particle systems suggest
that the effective 1D atom-dimer and dimer-dimer interactions
are to a good approximation determined by simple analytical 
expressions. Implications for the description of quasi-1D Fermi gases
within strict 1D frameworks
are discussed. 
\end{abstract}


\maketitle
\section{Introduction}
\label{introduction}
Ultracold atomic and molecular
gases are considered nearly ideal model systems
since their confining
geometry, size and interaction strength can be varied with unprecedented
control~\cite{review}.
A key goal of ongoing research activities 
is to experimentally 
determine the complete phase diagram of 
cold atom systems~\cite{science}. 
The
successful demonstration of this task would provide a first step
towards 
utilizing cold atom systems as quantum 
emulators. 
The determination of phase diagrams of quasi-1D systems 
has received considerable attention
since 
these systems
can, under certain circumstances,
be described
by 1D model Hamiltonian whose properties have been studied 
extensively in the 
literature~\cite{1dexact}.
For this class of systems, the challange is
to establish which aspects of quasi-1D cold atom experiments 
can be described by
1D model Hamiltonian.

Naively, the 1D scattering strength between
two particles in a wave guide geometry may be estimated by integrating out the
tightly-confined 
transverse degrees of freedom. However, 
while the result is accurate
in the weakly-interacting regime, Olshanii's seminal work~\cite{olsh98} shows that 
the 1D scattering
strength $g_{\mathrm{1D}}^{\mathrm{aa}}$
depends in general non-trivially on
the 3D $s$-wave atom-atom scattering length $a_{\mathrm{3D}}^{\mathrm{aa}}$ and
the transverse angular frequency $\omega_{\rho}$.
The coupling 
constant $g_{\mathrm{1D}}^{\mathrm{aa}}$ determined by Olshanii is now 
widely used in many-body studies of 
Bose and
Fermi gases.
The applicability of an  effective atomic 1D Hamiltonian whose 
two-body interactions are
parameterized in terms of
$g_{\mathrm{1D}}^{\mathrm{aa}}$
has, e.g., been confirmed for a Bose gas
under highly elongated harmonic confinement
by comparing the results of 3D and 1D Monte Carlo calculations~\cite{astr04}.

Over the past few years, an
effective atomic 1D Hamiltonian has also been applied 
extensively
to two-component Fermi gases 
under highly-elongated 
confinement~\cite{1dpolarized,toka04fuch04,mora05};
in this case, however, the validity regime of the effective 
atomic 1D Hamiltonian has
not yet been assessed carefully.
It is clear that an effective atomic 1D Hamiltonian description breaks 
down when tightly-bound molecules form. In this case,
the system may be described by an effective 
molecular 1D Hamiltonian that treats each tightly-bound
molecule as a composite boson.
While the functional form of such an effective molecular Hamiltonian
is generally agreed upon, the parametrization of the effective 
atom-dimer and dimer-dimer interactions varies~\cite{toka04fuch04,mora05}.
Furthermore, it is not clear whether or not the effective atomic and molecular 
1D Hamiltonian descriptions connect smoothly in the strongly-interacting 
regime.

This work presents 3D and 1D zero-temperature {\em{ab initio}} calculations
for small two-component Fermi gases 
with up to $N=4$ atoms under highly-elongated confinement 
and assesses the validity regimes of effective atomic
and molecular 1D Hamiltonian.
Our main findings are:
{\em{i)}} The 3D energies are reproduced well by an 
effective atomic 1D
Hamiltonian for small $|a_{\mathrm{3D}}^{\mathrm{aa}}/a_{\rho}|$ 
($a_{\mathrm{3D}}^{\mathrm{aa}}<0$), where
$a_{\rho}$ denotes the oscillator length in the
tight confinement direction [see Eq.~(\ref{eq_ho})].
{\em{ii)}} For small positive
$a_{\mathrm{3D}}^{\mathrm{aa}}/a_{\rho}$,
the 3D energies are reproduced well by an effective molecular 1D 
Hamiltonian that depends on
the effective 1D atom-dimer and 
dimer-dimer scattering lengths $a_{\mathrm{1D}}^{\mathrm{ad}}$
and $a_{\mathrm{1D}}^{\mathrm{dd}}$;
analytical expressions 
for $a_{\mathrm{1D}}^{\mathrm{ad}}$
and $a_{\mathrm{1D}}^{\mathrm{dd}}$
are presented.
{\em{iii)}} 
For two of the energy curves considered (see below),
the descriptions based on the effective
atomic and molecular 1D Hamiltonian join
fairly smoothly
in the strongly-interacting regime, defined
through $|a_{\mathrm{3D}}^{\mathrm{aa}}| \gtrsim a_{\rho}$;
not surprisingly, the 
dependence of the energies
on the aspect ratio is largest in the strongly-interacting regime.

Our assessment of the validity regimes
of the effective atomic and molecular 1D Hamiltonian for small systems is 
expected to provide guidelines for larger systems, and is thus
of great importance
for realizing 
condensed matter and materials analogs as well as
for exploiting cold atom systems for
quantum computation and quantum simulation.
Quasi-1D few-fermion systems can be prepared by
loading a gas of ultracold fermions into an 
optical 
lattice~\cite{lattice}.
Measurements of the excitation spectrum
as a function of the interaction strength
would 
provide a stringent test of our microscopic predictions.

Section~\ref{model}
introduces the 3D model Hamiltonian, discusses the numerical
techniques employed to solve the corresponding Schr\"odinger equation
and presents the resulting 3D energies.
Section~\ref{results} introduces the effective atomic and molecular
1D Hamiltonian and presents detailed comparisons
between the 3D and 1D energies.
Section~\ref{excitation} discusses the excitation spectrum 
of strongly-interacting two-component
Fermi gases under highly-elongated
cylindrically-symmetric 
confinement.
Finally,
Sec.~\ref{conclusions} concludes.

\section{Full 3D Treatment: Energetics}
\label{model}
This section introduces the 3D model Hamiltonian and 
the numerical techniques employed to solve the corresponding Schr\"odinger
equation.
3D energies 
are presented for $N=2-4$ fermions under highly-elongated
confinement.

\begin{figure}
\vspace*{.2cm}
\includegraphics[angle=0,width=55mm]{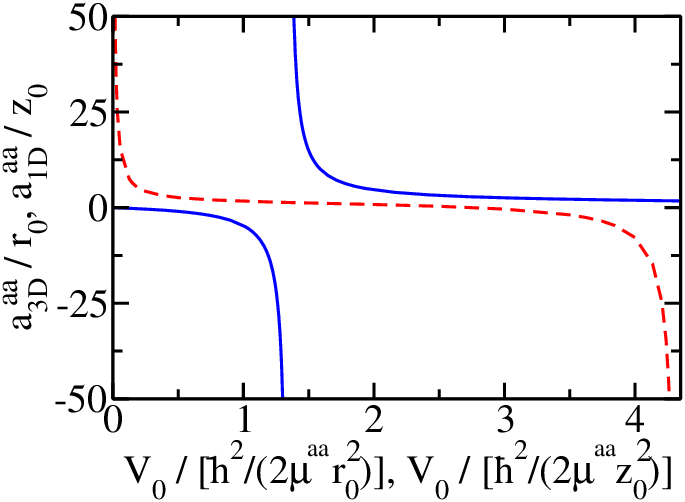}
\caption{
(Color online)
The solid and dashed lines show the free-space scattering lengths 
$a_{\mathrm{3D}}^{\mathrm{aa}}$ and 
$a_{\mathrm{1D}}^{\mathrm{aa}}$ for the Gaussian 
potential $V_{\mathrm{tb}}$
as a function of the well depth $V_0$.
The scattering lengths and depths are measured in the ``natural 
units'' of the free-space system, i.e., in units of $r_0$ [$z_0$]
and $\hbar^2/(2 \mu^{\mathrm{aa}} r_0^2)$ 
[$\hbar^2/(2 \mu^{\mathrm{aa}} z_0^2)$] for the
3D [1D] system, where $\mu^{\mathrm{aa}}$ denotes the reduced
mass of the atom-atom system.
}\label{ascatt}
\end{figure}

Our 3D model Hamiltonian $H_{\mathrm{3D}}$ 
for the trapped two-component Fermi gas with 
$N_1$ 
spin-up and $N_2$ spin-down 
fermions, where $N=N_1+N_2$, reads
\begin{eqnarray}
\label{eq_ham}
H_{\mathrm{3D}} = 
\sum_{i=1}^{N} \left[
\frac{-\hbar^2}{2m} \nabla^2_{\vec{r}_i} + 
V_{\mathrm{tr}}(\vec{r}_i)
\right] + 
\sum_{i=1}^{N_1} \sum_{j=N_1+1}^{N} 
V_{\mathrm{tb}}({r}_{ij}).
\end{eqnarray}
Here, $m$ and
$\vec{r}_i$ denote the atom mass and the position
vector of the $i$th atom,
$\vec{r}_i=(x_i,y_i,z_i)$.
The trapping potential $V_{\mathrm{tr}}(\vec{r}_i)$ is given by
\begin{eqnarray}
V_{\mathrm{tr}}(\vec{r}_i)=\frac{1}{2}
m \omega_z^2 (\lambda^2 \rho_i^2 + z_i^2),
\end{eqnarray}
where $\rho_i$ and $\lambda$ are defined through
$\rho_i=\sqrt{x_i^2+y_i^2}$ and 
\begin{eqnarray}
\label{eq_aspect}
\omega_{\rho}=\lambda \omega_z,
\end{eqnarray}
and
$\omega_{\rho}$ and $\omega_z$ denote
the transverse and axial angular frequencies.
Unlike atoms interact through
a spherically symmetric short-range
Gaussian potential $V_{\mathrm{tb}}$, 
\begin{eqnarray}
\label{eq_gaussian}
V_{\mathrm{tb}}(r_{ij})=-V_0 
\exp \left(-\frac{r_{ij}^2}{2r_0^2}  \right),
\end{eqnarray} 
where
${r}_{ij} =  |\vec{r}_i - \vec{r}_{j}|$.
We take the range $r_0$ to be much smaller than the
oscillator lengths $a_{z}$ and $a_{\rho}$ in the 
$z$- and $\rho$-directions,
\begin{eqnarray}
\label{eq_ho}
a_{z,\rho}=\sqrt{\frac{\hbar}{m \omega_{z,\rho}}},
\end{eqnarray}
and adjust
the depth $V_0$ ($V_0>0$)
so that the free-space 
3D $s$-wave atom-atom scattering length $a_{\mathrm{3D}}^{\mathrm{aa}}$
takes the desired value.
A solid line in Fig.~\ref{ascatt} 
shows $a_{\mathrm{3D}}^{\mathrm{aa}}$ as a function of the
well depth $V_0$.
To realize different negative $a_{\mathrm{3D}}^{\mathrm{aa}}$, we start with a 
non-interacting (NI) system ($V_0=0$)
and increase the depth $V_0$
till $|a_{\mathrm{3D}}^{\mathrm{aa}}|$ becomes infinitely large;
at this point, the free-space two-particle system supports a single
zero-energy $s$-wave bound state. 
To realize different positive $a_{\mathrm{3D}}^{\mathrm{aa}}$, we
increase $V_0$ further.
In general, $V_{\mathrm{tb}}$ can lead not only
to $s$-wave scattering but also to higher partial wave scattering.
We have checked that the generalized $p$-wave scattering 
length and generalized
scattering lengths corresponding to other higher 
partial waves are negligible over the range of well depths considered 
in this paper.
This implies that 
$H_{\mathrm{3D}}$ effectively describes an $s$-wave interacting system.

To solve the time-independent Schr\"odinger equation for 
$H_{\mathrm{3D}}$,
we separate off the
center-of-mass motion and numerically solve the
resulting Schr\"odinger equation in the
relative coordinates. For 
$N_1=N_2=1$,
we expand the relative wave function in 
terms of two-dimensional B-splines 
and diagonalize the Hamiltonian matrix. 
For the three- and four-particle 
systems, we employ a correlated Gaussian (CG) 
approach~\cite{cgbook,cgother}
that expands the relative wave function $\psi$ in terms of
Gaussian basis functions $f_k^{(\rho)} f_k^{(z)}$,
\begin{eqnarray}
\label{eq_cg}
\psi = \sum_{k=1}^{N_b} c_k {\cal{A}} \nonumber \\
\left[
f_{k}^{(\rho)}(\rho_{12},\cdots,\rho_{N-1,N}) 
f_{k}^{(z)}(z_{12},\cdots,z_{N-1,N})
\right],
\end{eqnarray}
where
\begin{eqnarray}
\label{eq_cg2}
f_{k}^{(\rho)}(\rho_{12},\cdots,\rho_{N-1,N}) =
\exp \left[ -\sum_{i<j}^N 
\left( \frac{\rho_{ij}}{\sqrt{2} d_{ij,k}^{(\rho)}} \right) ^2 
\right]
\end{eqnarray}
and $f_k^{(z)}$ is defined analogously.
The relative 
coordinates $\rho_{ij}$
and $z_{ij}$ are defined as
$\rho_{ij}=\sqrt{(x_i-x_j)^2+(y_i-y_j)^2}$ and 
$z_{ij}=z_i-z_j$ ($i,j=1,\cdots,N$ with $i<j$).
The widths $d_{ij,k}^{(\rho)}$ 
and $d_{ij,k}^{(z)}$  are chosen semi-stochastically for each 
pair $ij$ and $k$th basis function, and
the total number of basis functions is denoted by $N_b$.
In Eq.~(\ref{eq_cg}),
the $c_k$ denote expansion coefficients
and 
${\cal{A}}$ denotes an anti-symmetrizer that ensures the proper symmetry
of the two-component Fermi gas
under exchange of identical fermions. 
For $N=3$ ($N_1=2$ and $N_2=1$), ${\cal{A}}$ can be 
conveniently written as $1-P_{12}$, where $P_{12}$ permutes
the two up-fermions.
For $N=4$
($N_1=N_2=2$), ${\cal{A}}$ can be written as
$1-P_{12}-P_{34} + P_{12}P_{34}$.

For the interaction and confining potentials chosen,
the Hamiltonian and overlap matrix elements (the basis 
functions $f_k^{(\rho)} f_k^{(z)}$ do not form an orthogonal
set)
can be constructed analytically.
The diagonalization of the eigenvalue
equation is then performed using standard techniques.
The resulting eigenenergies,
whose accuracy
can be systematically improved by increasing the number of basis
functions and by optimizing the widths $d_{ij,k}^{(\rho)}$ 
and $d_{ij,k}^{(z)}$ of the Gaussian functions,
 provide upper bounds to the exact eigenenergies.

The 2D functions $f_k^{(\rho)}$ are eigenfunctions of the 
$z$-component $L_z$ of the orbital angular 
momentum operator with eigenvalues $\hbar m_l$,
$m_l=0$~\cite{cgbook}, while 
the 1D functions $f_k^{(z)}$ have
even parity $P_z=+1$.
For the $N=4$ system, the energetically lowest-lying state  
has $m_l=0$ and $P_z=+1$ for all 
3D scattering lengths $a_{\mathrm{3D}}^{\mathrm{aa}}$
and the basis functions $f_k^{(\rho)} f_k^{(z)}$ defined
in and below Eq.~(\ref{eq_cg})
have the proper symmetry. 
The ground state of the NI $N=3$ system,
in contrast,
has $m_l=0$ and odd parity ($P_z=-1$), which cannot be
described by the basis functions
$f_k^{(\rho)}f_z^{(z)}$.
To describe  states with odd parity, 
we add a spectator atom that does not 
interact with the $N$-fermion system of interest; 
the energy of the NI spectator
atom follows from its $m_l$ quantum number and from its parity,
and is subtracted at the end of the calculation.
Since the basis functions of the 
$(N+1)$-system have even parity, the spectator
atom and the $N$-fermion system either both have even parity
or both have odd parity.
In the following,
we label our solutions by the parity $P_z$;
if a spectator atom is added for computational purposes,
we report the parity of the physical system of interest. 
Furthermore, since all energetically lowest-lying states
of two-component Fermi gases under highly-elongated confinement
have $m_l=0$, we frequently omit the $m_l$ label. 

\begin{figure}
\vspace*{0.2cm}
\includegraphics[angle=0,width=65mm]{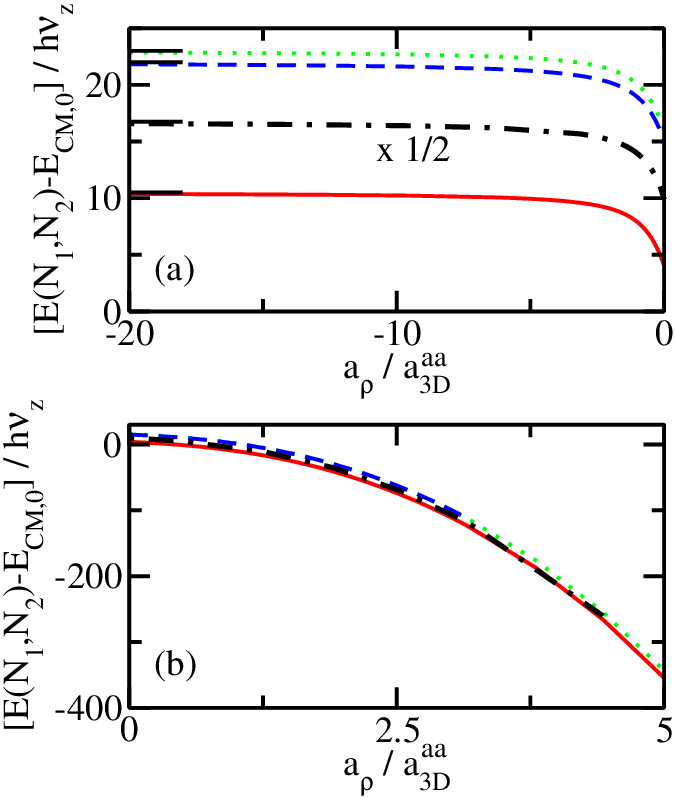}
\caption{
(Color online)
Relative 3D energies calculated using $H_{\mathrm{3D}}$
as a function of 
$a_{\rho}/a_{\mathrm{3D}}^{\mathrm{aa}}$ for (a)
negative
$a_{\mathrm{3D}}^{\mathrm{aa}}$
and (b) positive
$a_{\mathrm{3D}}^{\mathrm{aa}}$ for $\lambda=10$
[$r_0=0.03a_z$
for $a_{\mathrm{3D}}^{\mathrm{aa}}<0$ and 
$N=4$,
and $0.01a_z$ otherwise].
Solid lines show the $s$-wave energy $E_s(1,1)-E_{\mathrm{CM},0}$ of the
two-particle system, dashed lines show the lowest three-particle
energy $E(2,1)-E_{\mathrm{CM},0}$ with $P_z=-1$,
dotted lines show the lowest three-particle energy 
$E(2,1)-E_{\mathrm{CM},0}$ with $P_z=+1$
and
dash-dotted lines show one half of the lowest four-particle energy 
$E(2,2)-E_{\mathrm{CM},0}$ with $P_z=+1$.
The horizontal solid lines on the left side of panel~(a) indicate
the relative energies of the NI systems:
$E_s(1,1)-E_{\mathrm{CM},0}=10.5 \hbar \omega_z$,
$E(2,1)-E_{\mathrm{CM},0}=22 \hbar \omega_z$ ($P_z=-1$),
$E(2,1)-E_{\mathrm{CM},0}=23 \hbar \omega_z$ ($P_z=+1$), and
$[E(2,2)-E_{\mathrm{CM},0}]/2=16.75 \hbar \omega_z$.
}
\label{energyoverview}
\end{figure}

Figures~\ref{energyoverview}(a) and (b) show the 
relative 3D energies for $N=2-4$
and $\lambda=10$
as a function of the inverse 3D scattering length 
$a_{\rho}/a_{\mathrm{3D}}^{\mathrm{aa}}$.
Solid lines in Figs.~\ref{energyoverview}(a) and (b) show
the relative $s$-wave ground state energy $E_s(1,1)-E_{\mathrm{CM},0}$
of the two-body system.
For later convenience, the energy
$E_s(1,1)$ as well as the energies $E(N_1,N_2)$ (see below)
include the center-of-mass ground state energy
$E_{\mathrm{CM},0}$, 
$E_{\mathrm{CM},0}=\hbar \omega_{\rho} + \hbar \omega_z/2$.
In the NI limit, $E_s(1,1)-E_{\mathrm{CM},0}$
equals $10.5 \hbar \omega_z=1.05 \hbar \omega_{\rho}$.
In the absence of the confining potential in the $z$-direction, the
relative two-body energy is
always smaller than 
$\hbar \omega_{\rho}$, indicating the existence of a quasi-1D bound state for
all 3D scattering lengths 
$a_{\mathrm{3D}}^{\mathrm{aa}}$~\cite{olsh98,berg02}.
The confining potential in the
$z$-direction pushes the energy up; in the NI limit, the up-shift is 
given by
the zero-point energy $\hbar \omega_z/2$.

\begin{figure}
\vspace*{.2cm}
\includegraphics[angle=0,width=70mm]{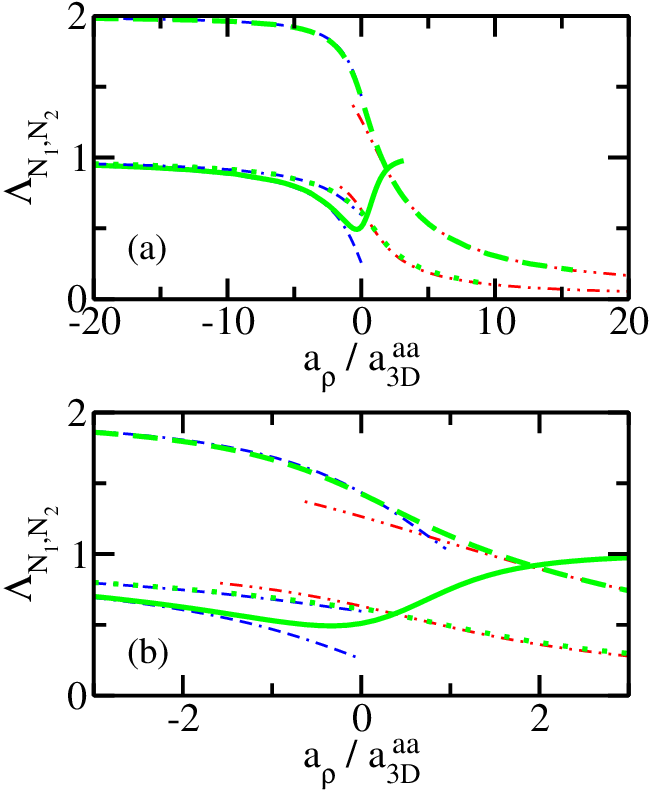}
\caption{
(Color online)
$\Lambda_{N_1,N_2}$ as a function of $a_{\rho}/a_{\mathrm{3D}}^{\mathrm{aa}}$:
Panel~(a) covers a large scattering length range while
panel~(b) shows an enlargement of the strongly-interacting regime.
Thick dotted, dashed and solid lines show 
$\Lambda_{2,2}$ 
($P_z=+1$),
$\Lambda_{2,1}$ 
($P_z=+1$) and
$\Lambda_{2,1}$ 
($P_z=-1$)
calculated using $H_{\mathrm{3D}}$ for $\lambda=10$
[$r_0=0.03a_z$
for $a_{\mathrm{3D}}^{\mathrm{aa}}<0$ and 
$N=4$,
and $0.01a_z$ otherwise].
Thin
dash-dash-dotted and dash-dot-dotted lines show the 
corresponding 1D energy curves 
calculated using $H_{\mathrm{1D}}^{\mathrm{a}}$ and
$H_{\mathrm{1D}}^{\mathrm{m}}$.}\label{fig2}
\end{figure}

For the $N=3$ system, the relative 
energies $E(2,1)-E_{\mathrm{CM},0}$ 
of the energetically lowest-lying states
with $P_z=+1$ and $-1$ are shown by dotted and dashed lines, 
respectively. 
The $P_z=-1$ state has lower 
energy for small $|a_{\mathrm{3D}}^{\mathrm{aa}}|$,
$a_{\mathrm{3D}}^{\mathrm{aa}}<0$ [see Fig.~\ref{energyoverview}(a)],
while the $P_z=+1$ state has lower energy for small positive
$a_{\mathrm{3D}}^{\mathrm{aa}}$ 
[the crossover of the two states 
is not visible on the
scale shown in Fig.~\ref{energyoverview}(b); it occurs at
$a_{\rho}/a_{\mathrm{3D}}^{\mathrm{aa}} \approx 2$
(see also Fig.~\ref{fig2})].
The relative
three-particle energies are just slightly larger than the relative 
two-body $s$-wave energies 
in the limit of 
small positive
$a_{\mathrm{3D}}^{\mathrm{aa}}$, indicating that the three-particle system 
can be thought of as consisting
of an 
$s$-wave dimer and an unpaired atom. The relative ground state energy 
$E(2,2)-E_{\mathrm{CM},0}$ of the
four-particle system
has $P_z=+1$ for all 3D scattering lengths
$a_{\mathrm{3D}}^{\mathrm{aa}}$;
to ease comparisons between the energies of the two- and
 four-particle systems,
dash-dotted lines in Figs.~\ref{energyoverview}(a) and (b) show one half 
of the
relative four-particle energy.
For
small 
positive $a_{\mathrm{3D}}^{\mathrm{aa}}$,
the four-particle energy approaches 
approximately twice the energy of the 
two-particle system, indicating that the 
four-particle system can be thought of as 
consisting of two $s$-wave molecules.
No 
tightly-bound
trimers or tetramers are formed
in the $a_{\mathrm{3D}}^{\mathrm{aa}} \rightarrow 0^+$ limit,
in agreement with results for zero-range 
interactions~\cite{petr03,petr04,moran3}.

The 3D energies 
can be combined to
define the universal energy curve
$\Lambda_{N_1,N_2}$
for a
two-component Fermi gas 
under external cylindrically symmetric confinement, 
\begin{eqnarray}
\label{eq_crossover}
\Lambda_{N_1,N_2}=
\frac{E(N_1,N_2) - N_{\mathrm{d}} E_s(1,1) - 
N_{\mathrm{f}} \hbar (\omega_{\rho} + \frac{1}{2} \omega_z)}
{E_{\mathrm{NI}}(N_1,N_2) - N \hbar (\omega_{\rho} + \frac{1}{2}\omega_z)},
\end{eqnarray}
where $N_{\mathrm{d}} = \min \{ N_1,N_2 \}$ and 
$N_{\mathrm{f}} = |N_1-N_2|$.
In Eq.~(\ref{eq_crossover}),
$E_{\mathrm{NI}}(N_1,N_2)$  denotes the energy of the NI system,
and
the energies
$E_{\mathrm{NI}}(N_1,N_2)$, $E(N_1,N_2)$ and $E_s(1,1)$ 
include the center-of-mass ground state
energy $E_{\mathrm{CM},0}$.
To remove dependencies of the total energy $E(N_1,N_2)$
of the trapped system 
on 
$V_{\mathrm{tb}}$,
the $s$-wave ground state energy $E_s(1,1)$ of the trapped two-particle system
is subtracted on the right hand side of Eq.~(\ref{eq_crossover}).
If $E(N_1,N_2)$ corresponds to the energetically lowest-lying
state of the NI system, the universal energy curve $\Lambda_{N_1,N_2}$
equals one.
Conversely, if $\Lambda_{N_1,N_2}$ equals zero 
in the $a_{\mathrm{3D}}^{\mathrm{aa}} \rightarrow 0^+$ 
limit, then this indicates that the 
system is effectively NI and that
induced interactions are absent.
The definition of the universal energy curve presented in 
Eq.~(\ref{eq_crossover}) for
cylindrically-symmetric two-component
Fermi systems constitutes a straightforward generalization of
that previously introduced for spherically-symmetric two-component 
systems~\cite{stec07b,stec08}.

Figure~\ref{fig2}
shows the 3D energy curves $\Lambda_{N_1,N_2}$
as a function of $a_{\rho}/a_{\mathrm{3D}}^{\mathrm{aa}}$
for $\lambda=10$; the energies used to calculate the $\Lambda_{N_1,N_2}$
are the same as those shown in Fig.~\ref{energyoverview}.
A thick dotted line 
shows 
$\Lambda_{2,2}$
calculated using the four-body energies $E(2,2)$ that
correspond to states with $m_l=0$ and $P_z=+1$.
The energy curve $\Lambda_{2,2}$ decreases monotonically
from 1 to approximately 0 as
$a_{\rho}/a_{\mathrm{3D}}^{\mathrm{aa}}$ increases 
from small negative to large positive values.
Thick dashed and solid lines in Fig.~\ref{fig2} show 
the energy curves $\Lambda_{2,1}$
for the energetically lowest-lying $N=3$ states
with $P_z=+1$ and $-1$, respectively.
It can be seen  that
these $N=3$ energy curves 
cross at
$a_{\rho}/a_{\mathrm{3D}}^{\mathrm{aa}} \approx 1.9$.
For $a_{\mathrm{3D}}^{\mathrm{aa}} \rightarrow 0^-$ (NI limit),
the ground state 
has $m_l=0$ and $P_z=-1$:
One spin-up and one spin-down atom occupy the ground state harmonic oscillator
orbital while the second up-atom occupies
the first excited state harmonic oscillator orbital.
For $a_{\mathrm{3D}}^{\mathrm{aa}} \rightarrow 0^+$,
in contrast, the ground state for
$N=3$
has $m_l=0$ and $P_z=+1$:
The 
system consists
of a tightly-bound dimer and an 
atom, 
which both
occupy the lowest trap state.
In 
this
limit,
the energy of the state with $P_z=-1$ is about $\hbar \omega_z$ larger than
the energy of the $P_z=+1$ state [note that
the energy difference is too small
to be visible on the scale shown in Fig.~\ref{energyoverview}(b)]. 
This suggests that the tightly-bound molecule and the unpaired atom
interact through effective 1D potentials that lead to even and odd parity 
scattering for $P_z=+1$ and $-1$, respectively (see 
also 
the next section).

\section{1D Treatment: Energetics and Effective 
Interactions}
\label{results}
This section considers effective atomic and molecular 1D Hamiltonian,
which 
assume that the
motion in the $\rho$-direction
is frozen, and compares the resulting 1D energies
with the 3D energies discussed in the previous section.
The applicability of the 1D model Hamiltonian and their parametrizations
are discussed in detail.

\begin{figure}
\vspace*{0.2cm}
\includegraphics[angle=0,width=70mm]{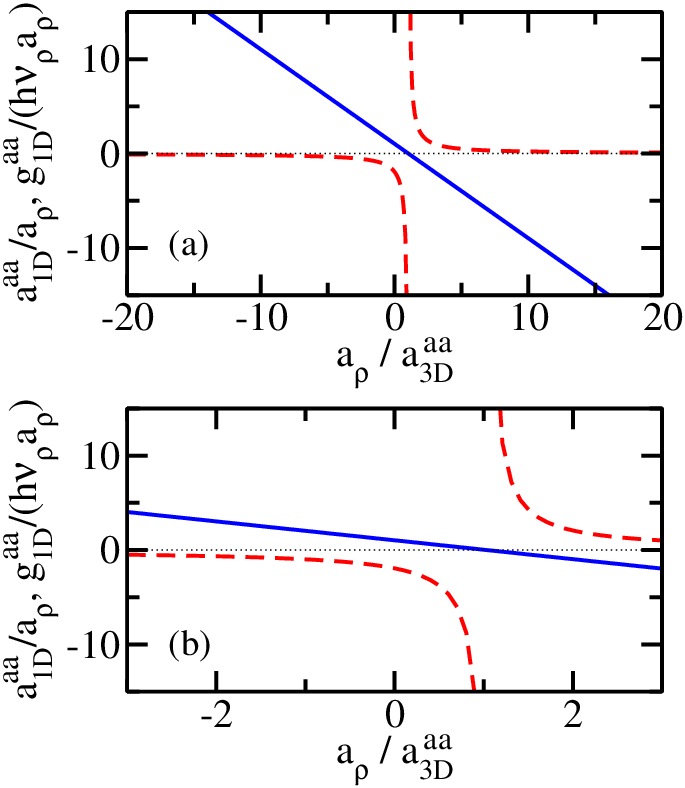}
\caption{
(Color online)
1D even parity atom-atom scattering length
$a_{\mathrm{1D}}^{\mathrm{aa}}/a_{\rho}$ (solid line) 
and corresponding 1D scattering strength 
$g_{\mathrm{1D}}^{\mathrm{aa}}/(\hbar \omega_{\rho} a_{\rho})$
(dashed line)
as a function of the inverse 3D atom-atom scattering length
$a_{\rho}/a_{\mathrm{3D}}^{\mathrm{aa}}$: 
Panel~(a) covers a large 3D scattering length range while
panel~(b) shows an enlargement of the strongly-interacting regime.
The 1D coupling constant $g_{\mathrm{1D}}^{\mathrm{aa}}$ changes from
about $-\hbar \omega_{\rho} a_{\rho}$ to
about $-60 \hbar \omega_{\rho} a_{\rho}$ as
$a_{\rho}/a_{\mathrm{3D}}^{\mathrm{aa}}$ increases from $-1$ to $1$ 
(i.e., in the
strongly-interacting regime).
}
\label{fig1}
\end{figure}

If the system behaves like an atomic gas,
the effective atomic 1D Hamiltonian 
$H_{\mathrm{1D}}^{\mathrm{a}}$ 
is given by 
\begin{eqnarray}
\label{eq_ham1d}
H_{\mathrm{1D}}^{\mathrm{a}} = 
\sum_{i=1}^{N} \left[
\frac{-\hbar^2}{2m} \frac{\partial^2}{\partial z_i^2} +
V_{\mathrm{tr}}(z_i)
\right] + 
\sum_{i=1}^{N_1} \sum_{j=N_1+1}^{N} 
V_{\mathrm{tb}}({z}_{ij}),
\end{eqnarray}
where 
\begin{eqnarray}
V_{\mathrm{tr}}(z_i)=\frac{1}{2}
m \omega_z^2  z_i^2.
\end{eqnarray}
In Eq.~(\ref{eq_ham1d}), the spin-up and spin-down fermions interact
through
the two-body potential 
$V_{\mathrm{tb}}(z_{ij})$ and, as in the 3D Hamiltonian $H_{\mathrm{3D}}$
[see Eq.~(\ref{eq_ham})], like atoms
do
not interact. 
The two-body potential $V_{\mathrm{tb}}$
is
chosen such that
its
1D even parity atom-atom scattering length $a_{\mathrm{1D}}^{\mathrm{aa}}$
is given by~\cite{olsh98}
\begin{eqnarray}
\label{eq_olshanii}
a_{\mathrm{1D}}^{\mathrm{aa}}= -\frac{\hbar}{2 \mu^{\mathrm{aa}} \omega_{\rho}} 
\left( \frac{1}{a_{\mathrm{3D}}^{\mathrm{aa}}} - 1.4603 \sqrt{\frac{\mu^{\mathrm{aa}} \omega_{\rho}}{\hbar}}  
\right),
\end{eqnarray}
where $\mu^{\mathrm{aa}}$ denotes the reduced mass of the atom-atom system.
The 1D even parity scattering length
$a_{\mathrm{1D}}^{\mathrm{aa}}$ [solid lines in Figs.~\ref{fig1}(a) and (b)]
 is large
for large $|a_{\rho}/a_{\mathrm{3D}}^{\mathrm{aa}}|$ 
($a_{\mathrm{3D}}^{\mathrm{aa}}<0$),
decreases linearly with increasing 
$a_{\rho}/a_{\mathrm{3D}}^{\mathrm{aa}}$, and
crosses zero 
at $a_{\rho}/a_{\mathrm{3D}}^{\mathrm{aa}}= 1.0326$.
Since the
1D coupling constant $g_{\mathrm{1D}}^{\mathrm{aa}}$,
\begin{eqnarray}
\label{eq_coupling}
g_{\mathrm{1D}}^{\mathrm{aa}}= -
\frac{\hbar^2}{\mu^{\mathrm{aa}}a_{\mathrm{1D}}^{\mathrm{aa}}},
\end{eqnarray}
diverges when $a_{\mathrm{1D}}^{\mathrm{aa}}$ vanishes, the quasi-1D system
is 
infinitely strongly-interacting for a finite 
$a_{\mathrm{3D}}^{\mathrm{aa}}$.
Furthermore, a large positive $a_{\mathrm{1D}}^{\mathrm{aa}}$ 
indicates the presence of a weakly-bound
even parity two-body
bound state.
In the literature, 
the effective 1D atom-atom potential
$V_{\mathrm{tb}}$ is frequently modeled by a 1D zero-range $\delta$-function
potential. 
For numerical convenience, we use instead
a 1D Gaussian potential [Eq.~(\ref{eq_gaussian}) with
$r_{ij}$ and $r_0$ replaced by $z_{ij}$ and $z_0$]
with a small width $z_0$
($z_0=0.005a_z$) and a depth adjusted so as 
to obtain the desired $a_{\mathrm{1D}}^{\mathrm{aa}}$.
We have checked that the resulting 1D energies 
depend only very weakly on $z_0$
and that the odd parity atom-atom
scattering length is negligibly small
over the range of well depths considered.
The 1D even parity atom-atom
scattering length $a_{\mathrm{1D}}^{\mathrm{aa}}$ 
for the 1D 
Gaussian potential is shown in Fig.~\ref{ascatt} 
by a dashed line
as a function of the well depth $V_0$.

For the 3D energy curves considered in Fig.~\ref{fig2},
the effective atomic 1D Hamiltonian is
expected to provide an accurate description 
if the size of the 1D dimer is much larger 
than the oscillator length $a_{\rho}$ (see, e.g., Ref.~\cite{toka04fuch04}).
Approximating the size of the dimer by 
$|a_{\mathrm{1D}}^{\mathrm{aa}}|$ and using 
$|a_{\mathrm{1D}}^{\mathrm{aa}}| \approx |a_{\rho}^2/a_{\mathrm{3D}}^{\mathrm{aa}}|$ 
[i.e., using the first part on the right hand side
of Eq.~(\ref{eq_olshanii})],
the validity condition 
$a_{\rho} \gg |a_{\mathrm{3D}}^{\mathrm{aa}}|$ is obtained.
Relaxing the disparity of length scales, we have  
$a_{\rho} \gtrsim |a_{\mathrm{3D}}^{\mathrm{aa}}|$ with 
$a_{\mathrm{3D}}^{\mathrm{aa}}<0$.

To obtain the 1D energies of the
effective atomic 1D Hamiltonian $H_{\mathrm{1D}}^{\mathrm{a}}$,
we 
first separate off the center-of-mass motion and then
solve the resulting 
Schr\"odinger equation in the relative coordinates using
the B-spline approach for the two-particle system
and 
the CG approach for the three- and four-particle systems. 
Our CG implementation 
for the 1D system parallels that 
discussed above for the 3D system. The main difference is that the basis
functions are 
now given by $f_k^{(z)}$ instead of by 
$f_k^{(\rho)} f_k^{(z)}$.
Systems with odd parity are, similarly to the 3D case, treated by adding
a NI spectator atom.
For large positive $a_{\mathrm{1D}}^{\mathrm{aa}}$, 
we find that the energetically lowest-lying 1D
states accurately model the corresponding 
3D states. For small positive 
$a_{\mathrm{1D}}^{\mathrm{aa}}$, however,
the effective atomic 1D Hamiltonian supports a sequence of tightly-bound
three- and four-particle states, which
have no analog in the 3D system (as discussed
above, tightly-bound three- and four-particle states are
not supported by $H_{\mathrm{3D}}$); these 1D states are
excluded from our analysis. The 1D energy states of interest
to us are those that smoothly evolve from a NI gas-like state
in the $a_{\mathrm{1D}}^{\mathrm{aa}} \rightarrow \infty$ limit
to states that describe 
a weakly-bound molecule and an atom or two weakly-bound molecules for
$N=3$ and $N=4$, respectively, in the 
$a_{\mathrm{1D}}^{\mathrm{aa}} \rightarrow 0^+$ limit.
Our last 
1D energies are reported for $a_{\rho}/a_{\mathrm{3D}}^{\mathrm{aa}}\approx 1$
for $N=3$ with $P_z=+1$, 
and for $a_{\rho}/a_{\mathrm{3D}}^{\mathrm{aa}}=0$ 
for $N=3$ with $P_z=-1$ and $N=4$ with $P_z=+1$.
We note that tightly-bound $N$-body states also exist for negative
$a_{\mathrm{1D}}^{\mathrm{aa}}$. Their
existence 
can be traced back to the finite range
of the Gaussian two-body
interaction potential; an effective atomic
1D Hamiltonian with zero-range $\delta$-function potentials
and negative $a_{\mathrm{1D}}^{\mathrm{aa}}$ 
does not support tightly-bound $N$-body states.

The 
1D energies 
determine the 1D energy curves 
$\Lambda_{N_1,N_2}$, which are given by Eq.~(\ref{eq_crossover})
with
$\omega_{\rho}=0$,
$E_s(1,1)$ denoting the 1D even 
parity two-particle
energy, and 
$E(N_1,N_2)$ and $E_{\mathrm{NI}}(N_1,N_2)$
interpreted as 1D energies.
Thin dash-dash-dotted lines in
Fig.~\ref{fig2}
show
the 1D energy curves obtained 
using $H_{\mathrm{1D}}^{\mathrm{a}}$
for 
$N=3$ and 4.
The agreement between the 1D energy curves and the corresponding 3D 
quantities
(thick lines) in the weakly-attractive regime is excellent.
For $P_z=+1$, the agreement between the 1D and 3D energy curves extends 
into the strongly-interacting, large positive 
$a_{\mathrm{3D}}^{\mathrm{aa}}$ regime.
The 1D energy curve for 
$N=3$
with $P_z=-1$, in contrast, starts deviating from the corresponding
3D energy curve for somewhat less strong interactions
($|a_{\rho}/a_{\mathrm{3D}}^{\mathrm{aa}}| \lesssim 2$
with $a_{\mathrm{3D}}^{\mathrm{aa}}<0$).
We have checked that these deviations are not due to 
the finite range of 
$V_{\mathrm{tb}}$.

In addition to an effective atomic 1D Hamiltonian 
$H_{\mathrm{1D}}^{\mathrm{a}}$,
we consider an effective molecular 1D Hamiltonian 
$H_{\mathrm{1D}}^{\mathrm{m}}$.
We show in the following that 
the 3D energy curves with $P_z=+1$
can be reproduced well
for $a_{\mathrm{3D}}^{\mathrm{aa}}>0$
by treating the
$N=3$ and 4
systems 
as effective two-particle systems that 
consist of an atom and a tightly-bound
molecule and of
two tightly-bound
molecules, respectively.
To this end, we model the 
atom-dimer and dimer-dimer interactions through a
$\delta$-function potential.
The effective two-particle 1D Hamiltonian for the relative 
coordinate $z$ then reads
\begin{eqnarray}
\label{eq_ham1dmolecule}
H_{\mathrm{1D}}^{\mathrm{m}} = \frac{-\hbar^2}{2 \mu^{j}} \frac{d^2}{dz^2} + 
\frac{1}{2} \mu^{j} \omega_z^2 z^2 
+g_{\mathrm{1D}}^{j}
\delta(z),
\end{eqnarray}
where 
$j={\mathrm{ad}}$ and $\mathrm{dd}$ for the atom-dimer and dimer-dimer
system, respectively, and where
the 1D coupling strengths $g_{\mathrm{1D}}^{\mathrm{ad}}$
and $g_{\mathrm{1D}}^{\mathrm{dd}}$ are related to 
the 1D scattering lengths $a_{\mathrm{1D}}^{\mathrm{ad}}$ and 
$a_{\mathrm{1D}}^{\mathrm{dd}}$
[Eq.~(\ref{eq_coupling})
with $\mathrm{aa}$ replaced by $\mathrm{ad}$ and 
$\mathrm{dd}$, respectively].
We 
approximate
the effective 1D atom-dimer and dimer-dimer scattering
lengths
$a_{\mathrm{1D}}^{\mathrm{ad}}$ and $a_{\mathrm{1D}}^{\mathrm{dd}}$
by the right hand side of
Eq.~(\ref{eq_olshanii}) with superscripts aa replaced by
ad and dd, respectively.
The 3D atom-dimer and dimer-dimer scattering lengths
$a_{\mathrm{3D}}^{\mathrm{ad}}$ and $a_{\mathrm{3D}}^{\mathrm{dd}}$, in turn,
are approximated by their free-space values~\cite{skor56,petr03,moran3,stec08},
\begin{eqnarray}
\label{eq_ad}
a_{\mathrm{3D}}^{\mathrm{ad}}=1.18 a_{\mathrm{3D}}^{\mathrm{aa}}
\end{eqnarray}
and~\cite{petr04,stec07b,stec08}
\begin{eqnarray}
\label{eq_dd}
a_{\mathrm{3D}}^{\mathrm{dd}}=0.608 a_{\mathrm{3D}}^{\mathrm{aa}}.
\end{eqnarray}
Physically, this implies that molecules are formed in 3D
and that their effective 3D interactions with atoms and other molecules 
are renormalized by the quasi-1D confinement.

The validity regime of the effective molecular 1D
Hamiltonian $H_{\mathrm{1D}}^{\mathrm{m}}$ is expected to be determined
by three conditions:
{\em{i}}) Since the 3D free-space atom-dimer and dimer-dimer scattering lengths
are derived assuming that $a_{\mathrm{3D}}^{\mathrm{aa}} \gg r_0$,
the above parametrization is expected to break down
when $a_{\mathrm{3D}}^{\mathrm{aa}}$ approaches $r_0$.
{\em{ii}}) For three- and four-particle
systems under spherically symmetric confinement, it
has been shown~\cite{stec08} that the full 3D 
energies on the BEC side (positive $a_{\mathrm{3D}}^{\mathrm{aa}}$)
are well described by effective 3D atom-molecule and molecule-molecule
models if $a_{\mathrm{3D}}^{\mathrm{aa}}$ is 
much smaller than the harmonic oscillator
length.
Correspondingly,
since our parametrization of the effective interactions
given in Eqs.~(\ref{eq_ham1dmolecule})-(\ref{eq_dd})
for the highly-elongated system
assumes that the molecules are formed
in 3D, the validity regime of $H_{\mathrm{1D}}^{\mathrm{m}}$ 
is 
expected to be given by
$a_{\mathrm{3D}}^{\mathrm{aa}} \ll a_{\rho}$.
{\em{iii}}) The effective 1D model treats the dimer as a point particle. This
treatment is justified if the atom-dimer and dimer-dimer distances are 
much larger 
than the size of the dimer,
i.e., if $a_z \gg a_{\mathrm{3D}}^{\mathrm{aa}}$ 
(see, e.g., Ref.~\cite{toka04fuch04}).
Combining the three criteria, we find that 
$H_{\mathrm{1D}}^{\mathrm{m}}$ is expected to
provide an accurate description if
$a_{\rho} \gg a_{\mathrm{3D}}^{\mathrm{aa}} \gg r_0$ or, 
employing less stringent criteria,
if $a_{\rho} \gtrsim a_{\mathrm{3D}}^{\mathrm{aa}} \gtrsim r_0$.
Combining this with the expected validity regime of
$H_{\mathrm{1D}}^{\mathrm{a}}$ (see above), the strongly-interacting
regime is defined through 
$|a_{\mathrm{3D}}^{\mathrm{aa}}| \gtrsim a_{\rho}$.
If the exact effective 1D atom-dimer and dimer-dimer scattering lengths
were known,
condition {\em{ii}}) would not apply and the 
expected validity regime of the
molecular 1D Hamiltonian would be larger 
($a_{z} \gtrsim a_{\mathrm{3D}}^{\mathrm{aa}} \gtrsim r_0$).

The Hamiltonian $H_{\mathrm{1D}}^{\mathrm{m}}$
given in Eq.~(\ref{eq_ham1dmolecule}) parametrizes the
effective interactions through a $\delta$-function potential
and thus assumes that the effective 1D atom-dimer and dimer-dimer
interactions lead to even parity scattering. Consequently,
the Hamiltonian
$H_{\mathrm{1D}}^{\mathrm{m}}$ 
does not describe the $P_z=-1$ energy curve for $N=3$.   
An effective molecular 1D model for the $N=3$ system with $P_z=-1$ 
would include an effective 1D interaction that leads to odd parity scattering
such as a so-called zero-range 
$\delta'$-potential~\cite{deltaprime}.
Although interesting, an effective molecular 1D description of the 
energy curve with $P_z=-1$ is not pursued
in this work.

Calculating the eigenenergies
of $H_{\mathrm{1D}}^{\mathrm{m}}$
from the known quantization condition~\cite{busc98}, we find that
the energy of the energetically lowest-lying 
state with gas-like character agrees well
with the 
3D quantities $E(2,1)-E_s(1,1)-\hbar\omega_{\rho}$
and
$E(2,2)-2E_s(1,1)$ for the atom-dimer and dimer-dimer
systems with $P_z=+1$ and $a_{\mathrm{3D}}^{\mathrm{aa}} \gtrsim 0$.
Dash-dot-dotted lines in Fig.~\ref{fig2}
show the energy curves for $P_z=+1$ calculated using the effective 
1D molecule model.
These 1D energy curves 
agree
well with the 
corresponding 3D energy curves
in the weakly-interacting regime.
Deviations are visible for
$N=3$ for 
$a_{\rho}/a_{\mathrm{3D}}^{\mathrm{aa}} \lesssim 1$
and for $N=4$ for 
$a_{\rho}/a_{\mathrm{3D}}^{\mathrm{aa}} \lesssim 0$.
The agreement of the 1D and 3D energy curves over a wide
range of interaction strengths
{\em{a posteriori}} justifies our parameterization
of the effective 1D atom-dimer and dimer-dimer scattering lengths
(see also Sec.~\ref{excitation}),
which differs from that employed in
earlier work~\cite{toka04fuch04,mora05,moran3}.
Notably, the 1D energy curves for the effective molecular
1D Hamiltonian connect nearly smoothly with those for the effective
atomic 1D Hamiltonian in the strongly-interacting regime.

\section{Excitation Spectrum}
\label{excitation}
This section discusses the behavior of the excitation frequency
$\omega_0$ for systems with
$P_z=+1$.
Within our 3D framework,
the excitation energy $\hbar \omega_0$
is defined as the difference between
the first excited and the lowest $P_z=+1$ states.
The corresponding 1D excitation energy is defined as the difference
between the energies of the corresponding 1D states.

\begin{figure}
\vspace*{.4cm}
\includegraphics[angle=0,width=70mm]{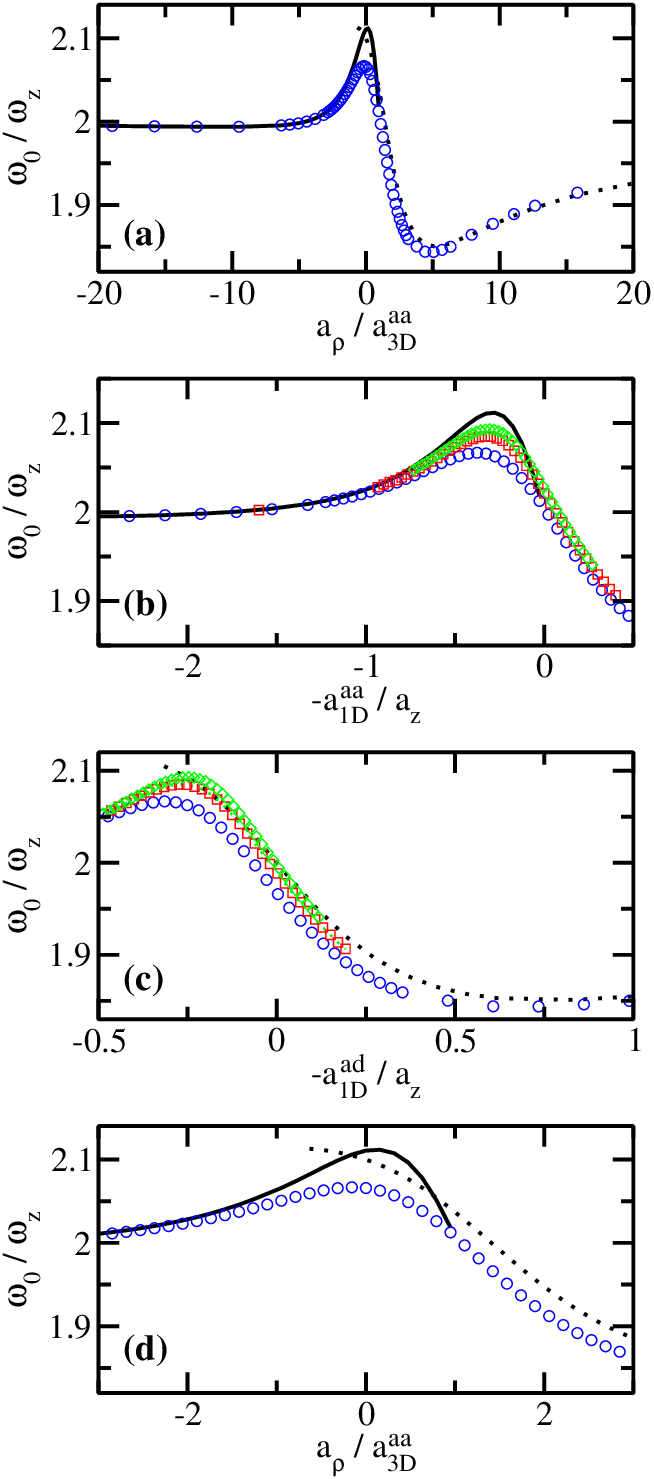}
\caption{
(Color online)
Excitation frequency $\omega_0/\omega_z$
for $N=3$ with
$P_z=+1$ and $r_0=0.01a_z$. 
(a) 
$\omega_0/\omega_z$ as a function of 
$a_{\rho}/a_{\mathrm{3D}}^{\mathrm{aa}}$ for $\lambda=10$. 
Circles show the 3D excitation frequency, while solid and dotted lines show
the corresponding 1D quantities calculated using 
$H_{\mathrm{1D}}^{\mathrm{a}}$ and $H_{\mathrm{1D}}^{\mathrm{m}}$,
respectively.
Panels~(b), (c) and (d) show the strongly-interacting regime 
in more detail.
(b) 
$\omega_0/\omega_z$ as a function 
of $-a_{\mathrm{1D}}^{\mathrm{aa}}/a_z$.
Circles, squares and diamonds show the 3D excitation frequency
for $\lambda=10$, $15$ and 20.
A solid line shows
the corresponding 1D quantities calculated using 
$H_{\mathrm{1D}}^{\mathrm{a}}$.
(c) $\omega_0/\omega_z$ as a function of 
$-a_{\mathrm{1D}}^{\mathrm{ad}}/a_z$.
Circles, squares and diamonds show the 3D excitation frequency
for $\lambda=10$, $15$ and 20.
A dotted line shows
the corresponding 1D quantities calculated using 
$H_{\mathrm{1D}}^{\mathrm{m}}$.
(d)
Enlargement of panel (a).
}
\label{fig4}
\end{figure}

Circles in Fig.~\ref{fig4}
show
the 3D excitation frequency
$\omega_0$ 
for $N=3$
with $P_z=+1$ and $\lambda=10$.
Panels~(a) and (d) show
$\omega_0$
as a function
of $a_{\rho}/a_{\mathrm{3D}}^{\mathrm{aa}}$.
The excitation frequency
$\omega_0$ equals $2 \omega_z$ 
in the NI limit ($a_{\mathrm{3D}}^{\mathrm{aa}}\rightarrow 0^-$),
reaches its maximum for infinitely large $a_{\mathrm{3D}}^{\mathrm{aa}}$ 
and
its minimum for $a_{\rho}/a_{\mathrm{3D}}^{\mathrm{aa}} \approx 5$,
and increases monotonically
towards $2 \omega_z$ as $1/a_{\mathrm{3D}}^{\mathrm{aa}}$ increases further.
To illustrate the dependence of $\omega_0$ on the
aspect ratio, squares and diamonds in Figs.~\ref{fig4}(b) and (c)
show the 3D
excitation frequency $\omega_0$ for two larger
aspect ratios, i.e., for $\lambda=15$ and 20.
Small dependencies of $\omega_0$ 
on $\lambda$ are visible in the strongly-interacting 
regime.

To ease comparisons 
between the 3D excitation frequencies
and those  based on the 1D Hamiltonian,
Figs.~\ref{fig4}(b) and (c) show enlargements of 
the strongly-interacting regime as functions of 
$-a_{\mathrm{1D}}^{\mathrm{aa}}/a_z$ and $-a_{\mathrm{1D}}^{\mathrm{ad}}/a_z$.
These scales are chosen since 
$a_{\mathrm{1D}}^{\mathrm{aa}}$ and $a_{\mathrm{1D}}^{\mathrm{ad}}$
determine
the properties of $H_{\mathrm{1D}}^{\mathrm{a}}$ and 
$H_{\mathrm{1D}}^{\mathrm{m}}$, respectively.
Solid lines in Fig.~\ref{fig4} show the excitation
frequencies calculated using the effective
atomic 1D Hamiltonian.
These
1D
excitation frequencies
reproduce the 3D excitation frequencies well
in the weakly-attractive regime
($a_{\mathrm{1D}}^{\mathrm{aa}}$ large). 
In the strongly-interacting regime, the agreement improves 
with increasing $\lambda$ [see Fig.~\ref{fig4}(b)].
Dotted lines in Fig.~\ref{fig4} show the 1D excitation frequencies
calculated using $H_{\mathrm{1D}}^{\mathrm{m}}$.
For large 
$|a_{\mathrm{1D}}^{\mathrm{ad}}|$ ($a_{\mathrm{1D}}^{\mathrm{ad}}<0$),
the 3D excitation frequencies are independent of $\lambda$
and reproduced
well by the 1D molecular model.
For smaller $|a_{\mathrm{1D}}^{\mathrm{ad}}|$,
the agreement improves 
with increasing $\lambda$ [see Fig.~\ref{fig4}(c)].

The effective molecular 1D Hamiltonian $H_{\mathrm{1D}}^{\mathrm{m}}$
predicts that a subset of
the $P_z=+1$ energy spectrum of the three- and
four-particle 
systems coincides with that of a two-particle
Tonks-Girardeau (TG) gas
for 
$a_{\mathrm{1D}}^{\mathrm{ad}}=0$~\cite{olsh98}.
For this atom-dimer scattering length, 
the effective molecular 1D Hamiltonian
predicts $\Lambda_{2,1}=1$, $\Lambda_{2,2}=1/2$ and
$\omega_0=2 \omega_z$.
Assuming that the behavior of the effective
dimer system is indeed governed by
$a_{\mathrm{1D}}^{\mathrm{ad}}$ (i.e., 
assuming that effective range and other corrections are negligible),
the condition $a_{\mathrm{1D}}^{\mathrm{ad}}=0$ 
signals an atom-dimer resonance.
Our 3D calculations for the three-particle system with $P_z=+1$
show that 
the ground state energy corresponds to that of a TG gas for
$a_{\rho}/a_{\mathrm{3D}}^{\mathrm{aa}} \approx 1.54-1.58$ and that
$\omega_0$ equals $2 \omega_z$ for 
$a_{\rho}/a_{\mathrm{3D}}^{\mathrm{aa}} \approx 1.1-1.4$ for $\lambda=10-20$,
in fairly good agreement with the prediction based on the 1D model,
$a_{\rho}/a_{\mathrm{3D}}^{\mathrm{aa}}=1.4069$.
The good agreement between our 3D results and those based on
the effective molecular 1D Hamiltonian, which is based on 
a simple empirical parametrization of the effective
1D atom-dimer and dimer-dimer scattering lengths, 
suggests that the 
molecular 1D model
employed in this work
provides a viable
and fairly accurate description of the system. 

The
atom-dimer $s$-wave resonance of
quasi-1D
systems found here,
$a_{\rho}/a_{\mathrm{3D}}^{\mathrm{aa}} \approx 1.5$, 
is somewhat smaller than 
that found 
by 
Mora {\em{et al.}}~\cite{moran3}
by solving a set of integral
equations 
for
zero-range interactions,
$a_{\rho}/a_{\mathrm{3D}}^{\mathrm{aa}} \approx 1.85$~\cite{footnoteelongatedcombine}. 
The 3D Hamiltonian employed by Mora {\em{et al.}}
accounts for the same physics as our 3D Hamiltonian
and the determination of the effective 1D atom-dimer
scattering length should, at least in principle, be exact~\cite{moran3}.
It is not clear at present
why
our empirical molecular 1D Hamiltonian provides a seemingly better
description than Mora {\em{et al.}}'s results for 
$a_{\mathrm{1D}}^{\mathrm{ad}} \approx 0$.

We also analyzed the ground state energy and
excitation spectrum for $N=4$.
The four-particle 3D energies 
are harder to converge than the 
three-particle 3D energies, and comparisons between the full 3D excitation 
frequencies  and the corresponding 1D quantities 
are accompanied by non-neglegible
uncertainties. 
We find that
our 3D results are consistent with the dimer-dimer $s$-wave resonance 
value
predicted by 
$H_{\mathrm{1D}}^{\mathrm{m}}$ [Eq.~(\ref{eq_ham1dmolecule})
with $j=\mathrm{dd}$, 
and $a_{\mathrm{1D}}^{\mathrm{dd}}$ given by Eq.~(\ref{eq_olshanii})
with $\mathrm{aa}$ replaced by $\mathrm{dd}$],
$a_{\rho}/a_{\mathrm{3D}}^{\mathrm{aa}} = 0.89$~\cite{footnoteelongatedcombine}.

\section{Conclusions}
\label{conclusions}
In summary, we have presented
highly-accurate, microscopic 3D calculations 
for small highly-elongated Fermi gases
with $N=2-4$ atoms
and reported the energies as a function of the interaction strength,
covering the weakly-attractive and weakly-repulsive regimes as well
as the strongly-interacting regime.
In addition, the dependence of the 
energies on the aspect ratio 
was investigated for selected cases.
While the role of the aspect ratio is negligible in 
the weakly-interacting 
regimes, its role becomes more important in the strongly-interacting regime,
possibly indicating that vitual excitations of transverse modes
become relevant.
The full 3D
energy curves with $P_z=+1$ are reproduced to a good approximation
by effective atomic and molecular 1D models
whose effective interactions are 
given by simple 
analytical expressions that depend
on the atom-atom $s$-wave
scattering length $a_{\mathrm{3D}}^{\mathrm{aa}}$,
the aspect ratio $\lambda$ and the atom mass $m$.
We find that the energies obtained 
from these effective atomic and
molecular 1D Hamiltonian 
join fairly
smoothly in the strongly-interacting regime. 
Assuming that the effective 1D
atom-dimer and dimer-dimer
scattering lengths govern the behavior of the highly-elongated system,
we deduced the positions of confinement-induced atom-dimer and 
dimer-dimer resonances from our energies.
Whether the effective 1D models also connect fairly smoothly for
larger systems is a pressing questions, in particular
since the determination of the phase diagram of highly-elongated systems
often times relies
on strictly 1D treatments.

In the future, it will be interesting to extend the studies
presented here to larger population-balanced and
population-imbalanced two-component Fermi gases.
While some microscopic calculations exist for strictly 1D systems,
microscopic 3D treatments that accurately account for the dynamics
along the tight and loose confining directions
are challenging.
Furthermore, it will be 
interesting to compare the 1D 
results obtained here for small systems with those
obtained within the local density approximation 
and to extend analogous comparisons to larger 
systems.

Support by the NSF through
grant PHY-0555316
is gratefully acknowledged.


\begin{thebibliography}{10}
\bibitem{footnoteelongatedcombine}
{Eq.~(\protect\ref{eq_olshanii}) predicts an atom-dimer [dimer-dimer] resonance
  at $a_{\rho}/a_{\mathrm{3D}}^{\mathrm{aa}} =1.4603 \times \sqrt{2/3} \times
  1.18=1.4069$ [$a_{\rho}/a_{\mathrm{3D}}^{\mathrm{aa}} =1.4603 \times
  0.608=0.8879$], which is smaller [larger] than
  $a_{\rho}/a_{\mathrm{3D}}^{\mathrm{aa}}=- \zeta(1/2,2.2)/\sqrt{2}=1.8503$
  [$a_{\rho}/a_{\mathrm{3D}}^{\mathrm{aa}}=-
  \zeta(1/2,0.3)/\sqrt{2}=-0.0079$]~\protect\cite{moran3,mora05}.}
\end{thebibliography}
\end{document}